\documentclass[aps,twocolumn,showpacs,amssymb,%
floatfix,superscriptaddress]{revtex4}
\usepackage{graphicx}% Include figure files 
\usepackage{epsf}
\usepackage{dcolumn}% Align table columns on decimal point 
\usepackage{bm}% bold math
\usepackage{hyperref} 
\usepackage{latexsym}

\begin{document}
\def\av#1{\langle#1\rangle}
\def\etal{{\it et al\/.}}
\def\pc{p_{\rm c}}
\def\l{{\lambda}}
\def\hm{h_*}
\def\xm{x_*}
\def\remark#1{{\bf *** #1 ***}}

% Include your paper's title here

\title{Statistics of Cycles: How Loopy is your Network?} 
\author{Hern\'an D. Rozenfeld}
\affiliation{Department of Physics, Clarkson University,
Potsdam NY 13699-5820}
\author{Joseph E. Kirk} 
\affiliation{Department of Math and Computer Science, Clarkson University, Potsdam, NY
13699-5805}
\author{Erik M. Bollt}
\affiliation{Department of Math and Computer Science, Clarkson University, Potsdam, NY
13699-5805}
\affiliation{Department of Physics, Clarkson University,
Potsdam NY 13699-5820}
\author{Daniel ben-Avraham}
\affiliation{Department of Physics, Clarkson University,
Potsdam NY 13699-5820}

\begin{abstract}
We study the distribution of cycles of length $h$ in large networks (of size $N\gg1$) and find it to be an excellent ergodic estimator, even in the extreme inhomogeneous case of scale-free networks. The distribution is sharply peaked around a characteristic cycle length, $\hm\sim N^{\alpha}$.  Our results suggest that $\hm$ and the exponent $\alpha$ might usefully characterize broad families of networks.  In addition to an exact counting of cycles in hierarchical nets, we present a Monte-Carlo sampling algorithm for approximately locating $\hm$ and reliably determining $\alpha$.
Our empirical results indicate that for {\it small\/} random scale-free nets of degree exponent $\l$, $\alpha=1/(\l-1)$, and $\alpha$ grows as the nets become larger.
\end{abstract}
\pacs{89.75.Hc, 02.50.Cw, 05.40.$-$a, 0.50.$+$q} 
\maketitle

%\section*{Introduction}
Recently, there has been much interest in large networks arising in a natural or social context (the Internet
and the World Wide Web, networks of social contacts, networks of predator-prey, of flight connections, the power grid, etc.)~\cite{albert02,dorogovtsev02,bornholdt}.
Initially, such networks were believed to be modeled by Erd\H os-R\'enyi (ER) random graphs~\cite{ER} --- graphs obtained by realizing only a fraction $p$ of the $\case{1}{2}N(N-1)$ links that could potentially form between the $N$ nodes present.  Subsequently, Watts and Strogatz demonstrated that the neighbors of a node, in most of the networks in question, tend to be connected to one another as well.  This effect of {\em clustering}, absent in ER graphs, is neatly captured in their Small World network model~\cite{watts98,watts99}.  Then, B\'arabasi \etal,~\cite{barabasi,albert02} observed that the degree $k$ of nodes (number of links connected to a node) in realistic networks follows a power-law, or {\em scale-free\/} distribution: $P(k)\sim k^{-\l}$.
The scale-free property gives rise to exotic behavior of the networks, such as resilience to random dilution (the percolation transition does not take place for $\l<3$), on the one hand, and high vulnerability to removal of the most connected nodes, on the other hand, and has become a principal focus of attention~\cite{cohen}.

The importance ascribed to scale-free degree distributions often obscures the relevance of other attributes.  The question is whether there exist other global characteristics of nets, beside their degree
distribution, that are relevant to their performance (stability, ease of transport, searchability).
Here we propose that the statistics of cycles seems particularly promising in this respect.  Cycles are relevant to propagation along the net, and their statistics exhibits a high degree of ergodicity (the results do not vary much from one node to the next).  We find that, in the thermodynamic limit of very large nets ($N\gg 1$) the distribution of cycles of length $h$ is sharply peaked around a characteristic cycle length $\hm\sim N^{\alpha}$.  Thus the distribution can be characterized by a single figure of merit --- the exponent $\alpha$, which we refer to as the {\em loopiness\/} exponent.  Generically $\alpha\leq1$, but we shall see that for many well known examples $\alpha=1$, while for {\it small\/} random scale-free nets of degree exponent $\l$ our preliminary results suggest that $\alpha=1/(\l-1)$, and $\alpha$ grows as the nets become larger.

The question of ergodicity is particularly difficult in scale-free graphs.  The highly connected nodes are responsible for many of the special properties attributed to these nets (lack of a percolation transition, rapid transport), yet the lower-degree nodes account for most of the nets' mass.  This skewness makes it a challenge to identify properties representative of the net as a whole.  
Consider, for example, the clustering index, defined as  $C_i=E_i/\frac{1}{2}k_i(k_i-1)$~\cite{watts98} ($k_i$ is the node's degree, or number of neighbors, and $E_i$  is the number of edges connecting between those neighbors).  The overall clustering index, $C=\av{C_i}$, averaged over all the nodes of the net, is a commonly cited statistics:  In some scale-free networks $C$ can be orders of magnitude larger than the corresponding ER graphs (ER graphs with the same numbers of nodes and links)~\cite{albert02}.  However, the clustering index of highly connected nodes tends to be quite smaller than for nodes of small degree, and this variation is overlooked in the global average.  

The ergodicity problem is solved in the statistics of cycles in the following sense.  An $h$-cycle is a closed path through  $h$  connected links that is self-avoiding (does not revisit nodes, other than the first)~\cite{bollobas}.
Define the {\em global} statistics $N_h$ as the total number of distinct $h$-cycles in the graph (cyclic permutations of the nodes do not count).  The {\em local} counterpart, $N_h^{(i)}$, is the number of $h$-cycles that pass through node $i$.  We argue that in scale-free graphs, it is likely that any cycle of moderate length $h$ will pass through the most connected node, so the difference between $N_h$ and $N_h^{(i)}$ could be quite small, making the $N_h$ a good global statistics.

\begin{figure}
\includegraphics*[width=0.4\textwidth]{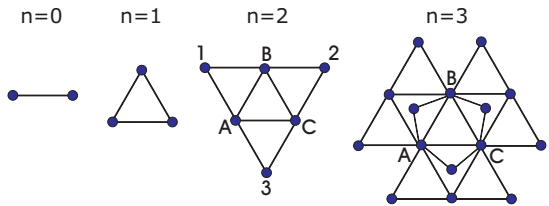}
\caption{Recursive scale-free graph, with $\l=1+\ln3/\ln2$.  Generation $n+1$ is obtained by joining three replicas at the hubs (most connected nodes) A, B, and C.  The closely related Sierpinski Gasket calls for joining the replicas at the vertices, 1, 2, and 3.}
\label{LoopyGraph}
\end{figure}

Consider the deterministic scale-free graph of Fig.~\ref{LoopyGraph}~\cite{pseudo}.  Each successive generation is obtained from the previous one by connecting a new node to both endpoints of every existing link.  Alternatively, the $(n+1)$-th generation could be constructed by adjoining three copies of generation $n$ at the hubs --- the most connected nodes (denoted by A, B, C, in Fig.~\ref{LoopyGraph}).   It can be shown that the degree of the nodes is distributed in scale-free fashion, with $\l=1+\ln3/\ln2$~\cite{pseudo}.
The recursive nature of this graph allows us to obtain the exact statistics of cycles.  Let $N_h(n)$ and $L_h(n)$ be the number of $h$-cycles, and the number of self-avoiding paths of length $h$ connecting between two hubs (A and B), in graphs of generation $n$, respectively.  Then,
\begin{equation}
\label{N(n)}
N_h(n+1)=3N_h(n)
+\!\!\!\!\!\!\!\!\!\!\!\sum_{{h_1,h_2,h_3\atop h_1+h_2+h_3=h}}\!\!\!\!\!\!\!\!\!\!\!L_{h_1}(n)L_{h_2}(n)L_{h_3}(n)\;,
\end{equation}
\begin{equation}
\label{L(n)}
L_h(n+1)=L_h(n)
+\!\!\!\!\!\sum_{{h_1,h_2\atop h_1+h_2=h}}\!\!\!\!\!\!L_{h_1}(n)L_{h_2}(n)\;,
\end{equation}
$N_h$ given by these relations is plotted in Fig.~\ref{Nh}.  Similar relations hold for the number of $h$-cycles that pass through a hub.  The two statistics become virtually identical beyond a small threshold $h$, confirming that $N_h$ does indeed constitute a good global estimator (Fig.~\ref{Nhhub}).

Evident in Fig.~\ref{Nh} is the scaling property 
\begin{equation}
\label{scaling}
\ln N_h=\hm f\big(\frac{h}{\hm}\big)\;, \qquad\hm\sim 2^n\;, 
\end{equation}
where the scaling function $f(x)$ is well-approximated by a parabola about its peak at $x=1$ (or $h=\hm$)~\cite{remark}.
Thus, the distribution of cycles, expressed in terms of the scaling parameter $x=h/\hm$, is nearly a Gaussian of width $1/\sqrt{\hm}$, and converges to a delta function in the thermodynamic limit of $\hm\to\infty$, or $N\to\infty$ (Fig.~{\ref{Nh}, inset).  It follows that in the thermodynamic limit the statistics of cycles is characterized by a single parameter, $\hm$, or better yet, by the way this quantity depends on the size of the net.  For the network of Fig.~\ref{LoopyGraph}, we have 
\begin{equation}
\label{lambda-1}
\hm\sim N^{\alpha},\qquad \alpha=\frac{\ln2}{\ln3}=\frac{1}{\l-1},
\end{equation}
(since the number of nodes in the net is $N(n)=(3^n+3)/2$, and $\hm(n)\sim2^n$).
We have tested other recursive nets and found various model-dependent exponents $\alpha\leq1$.
An interesting question which we examine below is what is $\alpha$ for {\it random\/} scale-free nets.

\begin{figure}
\includegraphics*[width=0.4\textwidth]{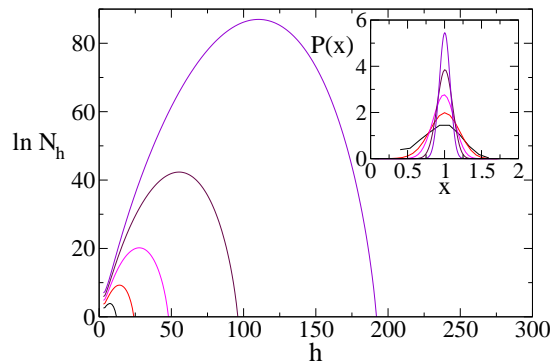}
\caption{Statistics of cycles for the network of Fig.~\ref{LoopyGraph}.  Shown are $\ln N_h$  for generations $n=3,4,5,6,7$.  Inset: The probability distribution for cycles of length $x=h/\hm$ tends to a delta function as $n\to\infty$ ($N\to\infty$). }
\label{Nh}
\end{figure}

\begin{figure}
\includegraphics*[width=0.4\textwidth]{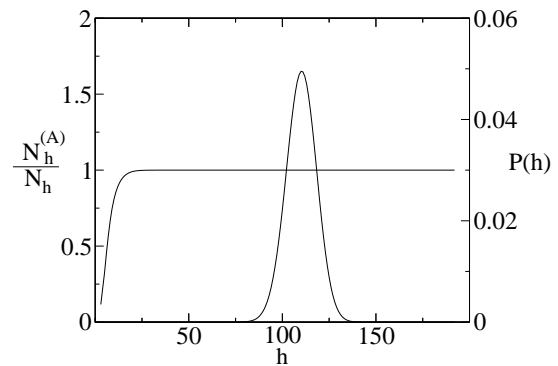}
\caption{The number of $h$-cycles that hit a hub, $N_h^{(A)}$, compared to the global statistics of $N_h$.  Shown are results for generation $n=7$, superposed upon a plot of $P(h)$.  Note the perfect agreement of the two statistics, where $P(h)$ is significant.}
\label{Nhhub}
\end{figure}

We now examine the statistics of cycles in other well known networks.  In a regular square lattice of $\sqrt{N}\times\sqrt{N}$ sites, Jensen and Guttmann find a similar scaling to that suggested by the example above, with $\hm\approx0.8N$, or $\alpha=1$~\cite{Jensen}.  It is also interesting to compare to the statistics of cycles in the Sierpinski Gasket, a fractal lattice which is closely related to the scale-free net of Fig.~\ref{LoopyGraph}.  For a given generation, the two graphs have the same number of nodes and links, but the Sierpinski Gasket constitutes almost a regular graph, where all nodes other than the three vertices share the same constant degree, $k=4$.  Here too, exploiting the recursive nature of the lattice, and following an exact counting procedure~\cite{sierpinski}, we find the same kind of scaling as in Eq.~(\ref{scaling}), but with $\hm\sim N$, or $\alpha=1$.

Next, consider a complete graph of order $N$, $K_N$.  Starting from an arbitrary node, the next node in
the cycle can be chosen from any of the remaining $N-1$ nodes, etc., yielding
\begin{equation}
\label{CompleteGraph}
N_h=\frac{N!}{2h(N-h)!}\;,\quad{\rm for\ complete\ graph}.
\end{equation}
The additional factor of $1/(2h)$ corrects for overcounting: it does not matter where a cycle starts ($h$ possibilities) and whether one traces it clockwise or counterclockwise.  At any rate, it follows that $\hm\approx N-1$ ($N\gg1$), and once again $\alpha=1$.  For the case of ER graphs, with only a fraction $p$ of the links realized, we cannot offer an exact expression but instead make the following approximation: Each link in an $h$-cycle is present with probability $p$, and so the whole cycle exists with probability $p^h$.  We then ignore the correlation between cycles (due to the fact that different cycles might share a subset of links) and write
\begin{equation}
\label{ERgraph}
N_h\approx\frac{N!}{2h(N-h)!}\,p^h\;,\quad{\rm for\ ER\ graphs}.
\end{equation}
This expression is of course exact in the limit of $p\to1$, and it correctly predicts the breakdown of cycles at the percolation transition threshold of $p_c=1/(N-1)$.  When $p$ is fixed and $N\to\infty$, we find once again $\hm\approx N$, suggesting $\alpha=1$.  If the $N\to\infty$ limit is approached along with $p=\omega/N\to0$, so as to keep next to the percolation transition, Eq.~(\ref{ERgraph}) predicts $\hm\approx(1-\omega^{-1})N$, and still $\alpha=1$.  However, the shortcomings of the approximation involved make this last result rather questionable.

It would seem that in most cases the self-avoiding cycles are space filling ($\alpha=1$), yet it is not clear whether this is true for ER graphs near the percolation transition.  At any rate, several recursive scale-free nets exhibit $\alpha<1$.  In order to study this and similar issues, we resort to a simple Monte-Carlo procedure for sampling the enormous number of cycles (of all sizes) that arise in various nets.

We first prune the net from all `dangling ends': nodes of degree $k=1$, and the link leading to the node, are removed from the net.  This action is reiterated until all extant nodes are of degree $k\geq 2$.
To find the frequency of cycles, we perform a self-avoiding random walk, starting from a randomly selected node.  Each step is chosen randomly between all the possibilities that would not result in self-intersection (other than with the starting node).  The walk is terminated when it comes back to the starting node, and the number of steps, $h$, is recorded.  A cycle produced in this way is a biased representative of the subset of $h$-cycles, because the excluded-volume constraint (the restriction of no self-intersection) is not uniform along the walk, becoming more severe as the walk progresses.  However, we know that this effect can be neglected in regular lattices of dimension $d\geq4$~\cite{degennes}.  We argue that the effect is likewise minimal in the environment of large, multiply connected networks.  

The frequency of $h$-cycles found out in this way is underestimated. Consider an arbitrary $h$-cycle already lain on the net.  Suppose that we are on node $i$ on the cycle and we take a step, choosing randomly from the $k_i-1$ links that would not force us back through the link leading to $i$.  The probability of hitting the next node on the cycle is $1/(k_i-1)$.  The probability of finding that particular
cycle is then proportional to $\Pi_i1/(k_i-1)$, where the product is taken over the nodes of the cycle. Thus, to better represent the true frequency of cycles, a cycle found by the self-avoiding random walk procedure is counted $\Pi_i(k_i-1)$ times.  This factor is actually too large, because some of the $k_i-1$ links, when followed from node $i$, lead only to dead ends (paths that self-intersect before completing the cycle).  The net effect is that the frequency of cycles of length $h$ is overestimated by an exponential factor $\sim e^{ch}$.  This is illustrated in Fig.~\ref{overestimate}, where we compare the sampled frequency of cycles, $P(h)'$, to their true frequency, $P(h)$, in a small ER graph of 18 nodes.  (The true frequency of cycles in such small nets can be counted by properly adapted depth-first-search algorithms.)

\begin{figure}
\bigskip
\includegraphics*[width=0.4\textwidth]{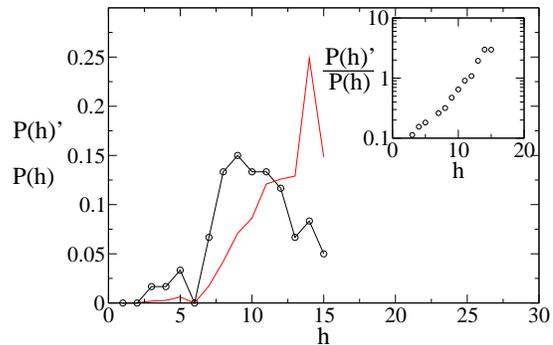}
\caption{Sampled frequency of cycles, $P(h)'$, compared to their true frequency ($\circ$), $P(h)$, in a small ER net of 18 nodes. Inset: The ratio $P(h)'/P(h)$ increases exponentially with $h$.}
\label{overestimate}
\end{figure}

As a result of the exponential overestimate, the most likely cycle is sampled at an apparent location: $\hm'=\hm+c\sigma^2/2$, where $\sigma$ is the width of the distribution $P(h)$.  In all cases examined above, $\sigma^2\sim N^{\alpha}$, so that $\hm'\sim N^{\alpha}$ and the sampling procedure yields the correct exponent $\alpha$.  (However, if $\sigma^2\sim N^{\beta}$, $\beta>\alpha$, our sampling procedure would find the exponent 
$\beta$ rather than $\alpha$.)  Indeed, when applied to the recursive nets of Fig.~\ref{LoopyGraph}, the sampling algorithm finds $\hm'\approx1.08\hm$, and correctly predicts $\alpha=0.63\pm0.02$ (compare with the exact result,
$\alpha=\ln2/\ln3\approx0.6309$).

\begin{figure}
%\bigskip
\includegraphics*[width=0.36\textwidth]{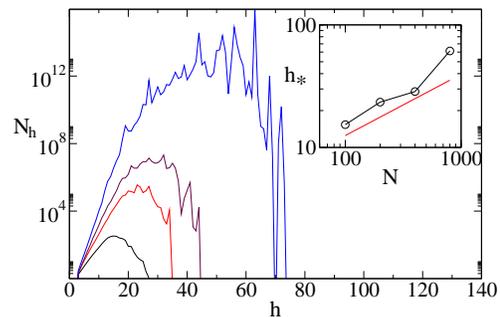}
\caption{Statistics of cycles in random scale-free graphs with $\l=3$.  Results from Monte-Carlo counting are shown for nets of size $N=100, 200, 400,800$.  Inset: Scaling of $\hm$ with $N$ ($\circ$) is consistent with $\hm\sim N^{\alpha}$, $\alpha=1/(\l-1)=0.5$ (solid line) only for $N$ small.}
\label{scalefree}
\end{figure}

We have applied the Monte-Carlo sampling to random scale-free graphs of degree exponent $\l=3$.
Our results, presented in 
Fig.~\ref{scalefree}, are consistent with the relation $\alpha=1/(\l-1)$ 
when the nets are small ($N\alt 200$), but $\alpha$ grows as $N$ increases.
This can be understood in the following way.  For $N\approx100$ and $\l=3$ we find that most cycles
are formed between the hub and nodes in the first shell (nodes connected to the hub by one link),
see Fig~\ref{smallSFnet}.
The few nodes in the second shell (two links away from the hub) that form part of cycles almost never
connect to one another.  The likely cycle length is then proportional to the number of nodes in the  first shell, or to the degree of the hub, $K\sim N^{1/(\l-1)}$~\cite{cohen,dor}.  Thus, for small $N$ the loop exponent is $\alpha=1/(\l-1)$.  As $N$ grows larger, nodes in
higher shells form part of interconnected cycles and $\alpha$ increases. 

\begin{figure}
%\bigskip
\includegraphics*[width=0.4\textwidth]{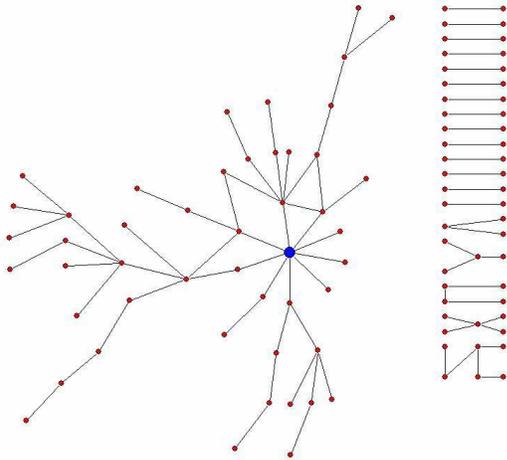}
\caption{Typical random scale-free net of $\l=3$ and $N=100$.  The net separates into
small, loopless components (shown on the right) and a giant component with loops involving
mostly the first shell of nodes attached to the hub.}
\label{smallSFnet}
\end{figure}

Cycles have been studied before, and several interesting results were obtained for cycles of small length, $h\ll N$~\cite{bollobas,cycles1,cycles2}.
Our study indicates that the {\em full\/} distribution of cycles, of all possible lengths, displays additional useful properties: Ergodicity is implied in the fact that the distribution of cycles that pass through a node is similar for most nodes of the net, even in the extreme inhomogeneous case of scale-free networks.  For large nets, the distribution resembles a delta function that peaks about a typical cycle size, $\hm\sim N^{\alpha}$.  The exponent $\alpha$ serves as a single figure of merit that characterizes the ``loopiness" of the net in question; the larger $\alpha$ the more loopy the net.  $\alpha=1$ for regular lattices and fractals, and for complete graphs.  For {\it small\/}, random scale-free nets $\alpha=1/(\l-1)$, but $\alpha$ increases as the nets become larger.  
It remains an open question whether the loopy exponent saturates at 
$\alpha=1$ as $N\to\infty$.

Among the many remaining open questions, finding reliable and efficient algorithms for sampling the distribution of cycles is perhaps the most important one. Only when these become available will we be able to study the full statistics of cycles in the truly large nets that have been the focus of so much recent attention.

\acknowledgments
We thank Larry M.~Glasser and Jos\'e F.~Mendes for many useful discussions.  We also thank the
referees for their critical input, forcing us to reevaluate our findings and conclusions regarding
random scale-free nets.  We are grateful to NSF grant PHY-0140094 (DbA) for partial support of this research.


\begin{thebibliography}{99}
\bibitem{remark} This scaling behavior can also be confirmed directly from Eqs.~(\ref{N(n)}), (\ref{L(n)}).
\end{thebibliography}
\end{document}